\documentclass[11pt]{article}
\usepackage{moriond}
\usepackage{minipage-marginpar}
\usepackage{epsfig}
\usepackage{graphicx}
\usepackage{graphics}
\usepackage{wrapfig}

\def\bea{\begin{eqnarray}}
\def\eea{\end{eqnarray}}
\def\beq{\begin{equation}}
\def\eeq{\end{equation}}

\newcommand{\ro}{\mbox{{\boldmath
$\rho$}}}

\newcommand{\pb}{\mbox{{\bf
p}}}

\newcommand{\bb}{{{\bf b}}}

\def\lsim{\mathrel{\rlap{\lower4pt\hbox{\hskip1pt$\sim$}}
    \raise1pt\hbox{$<$}}}         
\def\gsim{\mathrel{\rlap{\lower4pt\hbox{\hskip1pt$\sim$}}
    \raise1pt\hbox{$>$}}}         

\newcommand{\Photo}{}

\begin{document}
\vspace*{4cm}
\title{Jet tomography of $AA$-collisions at RHIC and LHC energies
}

\author{ B.G. ZAKHAROV }

\address{L.D. Landau Institute for Theoretical Physics,
        GSP-1, 117940,\\ Kosygina Str. 2, 117334 Moscow, Russia}

\maketitle\abstracts{
We present our recent results on jet tomography of AA-collisions
at RHIC and LHC. We focus on flavor dependence of the nuclear 
modification factor. The computations are performed accounting for 
radiative and collisional parton energy loss with running coupling 
constant.
}

\noindent{\bf 1.} 
In this talk I present results of jet tomographic analysis of the 
RHIC and LHC data on the nuclear modification factor 
$R_{AA}$ for light hadrons, single electrons, and $D$-mesons.
A major purpose of this study is to examine whether it is possible
in the pQCD picture 
of parton energy loss in the quark-gluon plasma (QGP) 
to describe simultaneously
quenching of light and heavy flavors.
One can expect that predictions for variation of 
$R_{AA}$ from light to heavy flavors should be more robust than 
that for $R_{AA}$ itself, which
have significant theoretical uncertainties.
The analysis is based on the
light-cone path integral approach \cite{LCPI,BSZ}. We evaluate $R_{AA}$
using the scheme developed in \cite{RAA08}.

\vspace{.2cm}
\noindent{\bf 2.}
We define the nuclear modification factor for a given impact parameter $b$
as
\beq
R_{AA}(b)=\frac{{dN(A+A\rightarrow h+X)}/{d\pb_{T}dy}}
{T_{AA}(b){d\sigma(N+N\rightarrow h+X)}/{d\pb_{T}dy}}\,,
\label{eq:10}
\eeq
where $\pb_{T}$ is the particle transverse momentum, $y$ is rapidity (we
consider the central region near $y=0$), 
$T_{AA}(b)=\int d\ro T_{A}(\ro) T_{A}(\ro-\bb)$, $T_{A}$ is the nucleus 
profile function. We write the differential yield for 
$A+A\to h+X$ process in  the numerator in the form
\beq
\frac{dN(A+A\rightarrow h+X)}{d\pb_{T} dy}=\int d\ro T_{A}(\ro)T_{A}(\ro-\bb)
\frac{d\sigma_{m}(N+N\rightarrow h+X)}{d\pb_{T} dy}\,,
\label{eq:20}
\eeq
where ${d\sigma_{m}(N+N\rightarrow h+X)}/{d\pb_{T} dy}$ is the medium-modified
cross section for the $N+N\rightarrow h+X$ process.
As in the ordinary pQCD formula, we write it as
\beq
\frac{d\sigma_{m}(N+N\rightarrow h+X)}{d\pb_{T} dy}=
\sum_{i}\int_{0}^{1} \frac{dz}{z^{2}}
D_{h/i}^{m}(z, Q)
\frac{d\sigma(N+N\rightarrow i+X)}{d\pb_{T}^{i} dy}\,,
\label{eq:30}
\eeq
where $\pb_{T}^{i}=\pb_{T}/z$ is the parton transverse momentum, 
${d\sigma(N+N\rightarrow i+X)}/{d\pb_{T}^{i} dy}$ is the
hard cross section,
$D_{h/i}^{m}$ is the medium-modified fragmentation function (FF)
for transition of a parton $i$ into the observed particle $h$.
For the parton virtuality scale $Q$ we take the parton transverse
momentum $p^{i}_{T}$.

In first approximation, overlap between the DGLAP and induced stages
of the parton showering can be neglected at $p_{T}\lsim 100$ GeV \cite{RAA08}.
Then, assuming that the final particle $h$ is formed outside the medium, 
the medium-modified FF can be written as
\beq
D_{h/i}^{m}(Q)\approx D_{h/j}(Q_{0})
\otimes D_{j/k}^{in}\otimes D_{k/i}(Q)\,.
\label{eq:40}
\eeq
Here $\otimes$ denotes $z$-convolution, 
$D_{k/i}$ is the ordinary DGLAP FF for $i\to k$ parton transition,
$D_{j/k}^{in}$ is the FF for $j\to k$ parton transition in the QGP
due to induced gluon emission, and 
$D_{h/j}$ describes fragmentation of the parton $j$ into
the detected particle $h$ outside of the QGP.

We computed the DGLAP FFs with the help of the PYTHIA event 
generator \cite{PYTHIA}.
For the stage outside the QGP for light partons
we use for $D_{h/j}(Q_{0})$ the 
KKP \cite{KKP} FFs  with $Q_{0}=2$ GeV.
We treat the formation of single electrons
from  heavy quarks as the two-step fragmentations
$c\to D\to e$ and $b\to B\to e$. 
For the $c\to D$ and $b\to B$ transitions we use 
the Peterson FF with parameters
$\epsilon_{c}=0.06$ and $\epsilon_{b}=0.006$.
The $z$-distributions for the $B/D\to e$ transitions 
have been calculated using the CLEO data 
\cite{CLEO_B,CLEO_D} on the electron spectra in the $B/D$-meson decays.
We did not include the $B\to D\to e$ process, which gives a negligible 
contribution. 

The one gluon induced spectrum has been computed using
the  method elaborated in \cite{Z04_RAA}. 
We take $m_{q}=300$ and $m_{g}=400$ MeV for the quark and gluon quasiparticle
masses,
for heavy quarks we take $m_{c}=1.2$ GeV,
and $m_{b}=4.75$ GeV. We use the Debye mass obtained 
in the lattice calculations \cite{Bielefeld_Md} that
give $\mu_{D}/T\sim 2.5\div 3$.
We use the running $\alpha_s$
frozen at some value $\alpha_{s}^{fr}$ at low momenta. For vacuum a
reasonable choice is $\alpha_{s}^{fr}\approx 0.7$ \cite{NZ_HERA}.
In plasma $\alpha_{s}$ can be reduced due to thermal effects,
and we regard $\alpha_{s}^{fr}$ as a free parameter of the model.
The multiple gluon emission has been
accounted for employing Landau's method
(for details see \cite{RAA08}).

We incorporate the collisional energy loss, 
which is relatively small \cite{Z_Ecoll}, by renormalizing
the initial temperature of the QGP, $T_{0}$, for the radiative FFs 
according to the following condition:
$\Delta E_{rad}(T^{\,'}_{0})=\Delta E_{rad}(T_{0})+\Delta E_{col}(T_{0})$, where
$\Delta E_{rad/col}$ is the radiative/collisional energy loss, $T_{0}$
is the real initial temperature of the QGP, and $T^{\,'}_{0}$ is the 
renormalized temperature. 
We calculate the collisional energy loss within 
Bjorken's method 
with an accurate treatment of kinematics of the $2\to 2$ 
processes (for details see \cite{Z_Ecoll}) with the same 
parametrization of $\alpha_{s}(Q)$ as for the radiative one.

We calculate the hard cross sections  using the LO 
pQCD formula with the CTEQ6 \cite{CTEQ6} PDFs.
To simulate the higher order $K$-factor
we take for the virtuality scale in $\alpha_{s}$ the value 
$cQ$ with $c=0.265$ as in the PYTHIA event generator \cite{PYTHIA}.
The nuclear modification of the parton densities
(which leads to some small deviation of $R_{AA}$ from unity even without
parton energy loss) has been incorporated with the help of the 
EKS98 correction \cite{EKS98}.

We describe the QGP in the Bjorken model 
with 1+1D expansion,  
which gives $T_{0}^{3}\tau_{0}=T^{3}\tau$. We take $\tau_{0}=0.5$ fm.
For simplicity 
we ignore variation of the initial temperature $T_{0}$ in the 
transverse directions in the overlapping of two nuclei.
We fix $T_{0}$ using 
the entropy/multiplicity ratio
$dS/dy{\Big/}dN_{ch}/d\eta\approx 7.67$ obtained in \cite{BM-entropy}.
It gives for central Au+Au collisions at $\sqrt{s}=200$ GeV
$T_{0}\approx 320$ MeV and for 
Pb+Pb collisions at $\sqrt{s}=2.76$ TeV
$T_{0}\approx 420$ MeV. 
The fast parton path length in the QGP, $L$, 
 has been calculated according to the geometry
of the hard process and $AA$-collision.
To account for the fact that at times about $1\div 2$ units of 
the nucleus radius the transverse expansion
should lead to a fast cooling of the hot QCD matter 
we impose the condition $L< L_{max}$. We take
$L_{max}=8$ ($L_{max}=10$ fm gives almost the same).
\begin{figure} [t]
\begin{center}
\epsfig{file=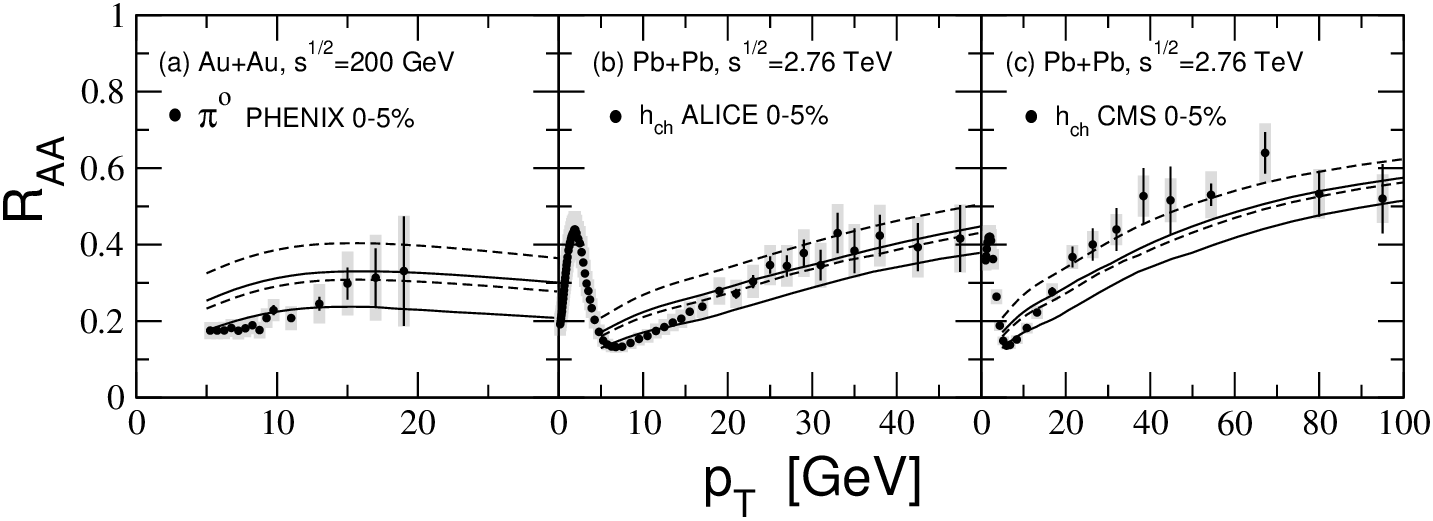,height=4.3cm,clip=,angle=0}
\end{center}
\caption[.]
{
(a) $R_{AA}$ for $\pi^{0}$ for 0--5\% central Au+Au collisions
at $\sqrt{s}=200$ GeV from our calculations compared to data from 
PHENIX \cite{PHENIX_RAA_pi}. 
(b,c) $R_{AA}$ for charged hadrons for 0--5\% central Pb+Pb collisions
at $\sqrt{s}=2.76$ TeV from our calculations compared to data from 
(b) ALICE \cite{ALICE_RAAch} and (c) CMS \cite{CMS_RAAch}.
Systematic experimental errors are shown as shaded areas. 
The curves show our calculations for radiative and collisional
energy loss (solid), and for purely radiative energy loss (dashed)
for $\alpha_{s}^{fr}=0.4$
(upper curves) and 0.5 (lower curves).
}
\end{figure}
\begin{figure}[tight]
\begin{minipage}[h]{0.49\linewidth}
\begin{center}
\includegraphics[width=0.76\linewidth,clip=]{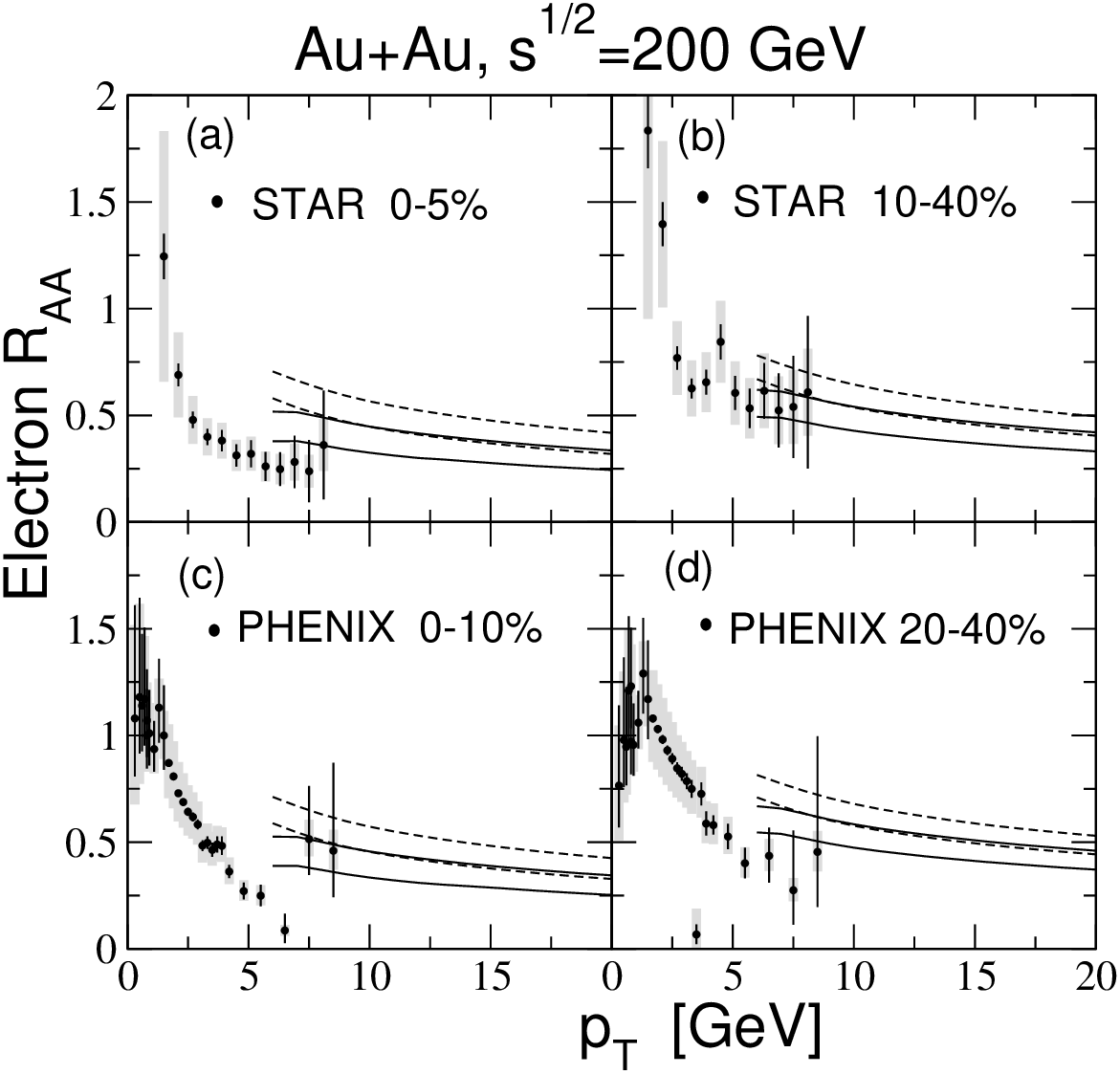}
\caption[.]
{
The electron $R_{AA}$ in Au+Au collisions at $\sqrt{s}=200$ 
GeV for (a) 0--5\%, (b) 10--40\%, (c) 0--10\%, (d) 20--40\% centrality
classes.  
The curves show calculations for radiative and collisional energy 
loss (solid), and for purely radiative energy loss (dashed) including
charm and bottom contributions for $\alpha_{s}^{fr}=0.4$
(upper curves) and 0.5 (lower curves).
Data points are from STAR \cite{STAR_e} and PHENIX \cite{PHENIX2_e}.
Systematic errors
are shown as shaded areas. 
}
\end{center}
\end{minipage}
\hfill
\begin{minipage}[h]{0.49\linewidth}
\begin{center}
\vspace{-9ex}
\includegraphics[width=0.77\linewidth,clip=]{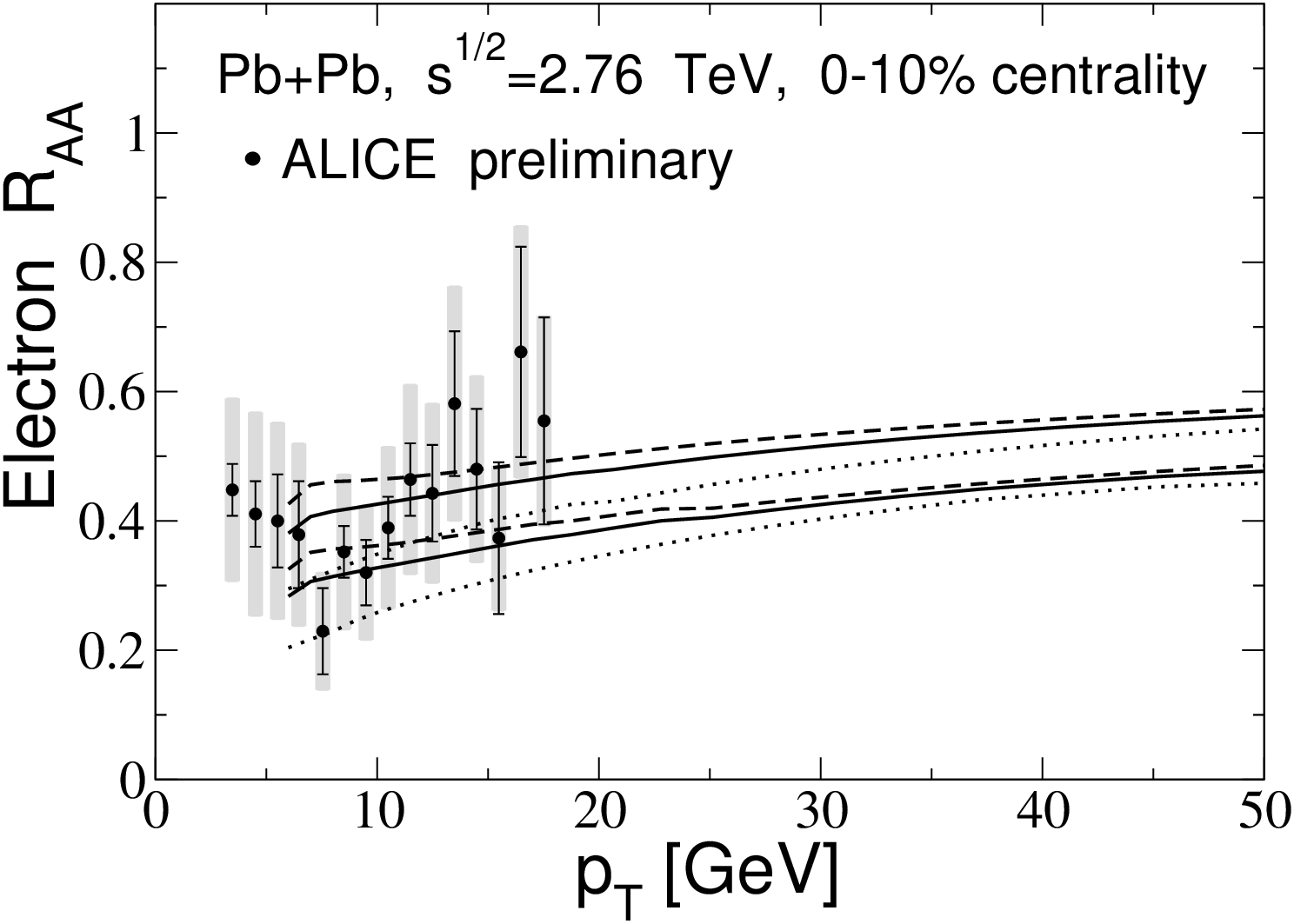}
\caption[.]
{
The electron $R_{AA}$ for
0--10\% central 
Pb+Pb collisions at $\sqrt{s}=2.76$ TeV for $\alpha_{s}^{fr}=0.4$
(upper curves) and 0.5 (lower curves).
The total $c+b\to e$  $R_{AA}$ (solid),
$c\to e$ (dashed), $b\to e$ (dotted) with collisional energy loss.
Data points are the preliminary ALICE data \cite{ALICE_e}.
Systematic errors
are shown as shaded areas. 
}
\end{center}
\end{minipage}
\end{figure}

\vspace{.2cm}
\noindent{\bf 3.}
Fig.~1 shows comparison of our predictions for $R_{AA}$ 
for $\alpha_{s}^{fr}=0.4$ and 
0.5 in 0--5\% centrality bin
for (a) $\pi^{0}$-mesons in  
Au+Au collisions at $\sqrt{s}=200$ GeV
to PHENIX data  \cite{PHENIX_RAA_pi},
and for (b,c) charged hadrons in 
Pb+Pb collisions at $\sqrt{s}=2.76$ TeV
to (b) ALICE \cite{ALICE_RAAch} and (c) CMS \cite{CMS_RAAch} 
data.
We show the total $R_{AA}$ with radiative and collisional energy loss
and for purely radiative energy loss.
One can see that the effect of the collisional mechanism
is relatively small (especially for LHC).
We present the results for $p_{T}\gsim 5$ GeV since for  smaller momenta 
our calculations of the induced gluon emission (based on the 
relativistic approximation) are hardly robust.
Fig.~1 shows that for light hadrons 
the window $\alpha_{s}^{fr}\sim 0.4\div 0.5$ 
leads to a reasonable magnitude of $R_{AA}$.
For RHIC the agreement 
of the theoretical $R_{AA}$
(radiative plus collisional energy loss) with the data  
is better for $\alpha_{s}^{fr}=0.5$. And for LHC 
the value $\alpha_{s}^{fr}=0.4$ seems to be preferred by the data
(if one considers the complete $p_{T}$ range).
\begin{wrapfigure}[11]{r}{0.5\linewidth}
\vspace{-15ex}
\includegraphics[width=\linewidth,clip=]{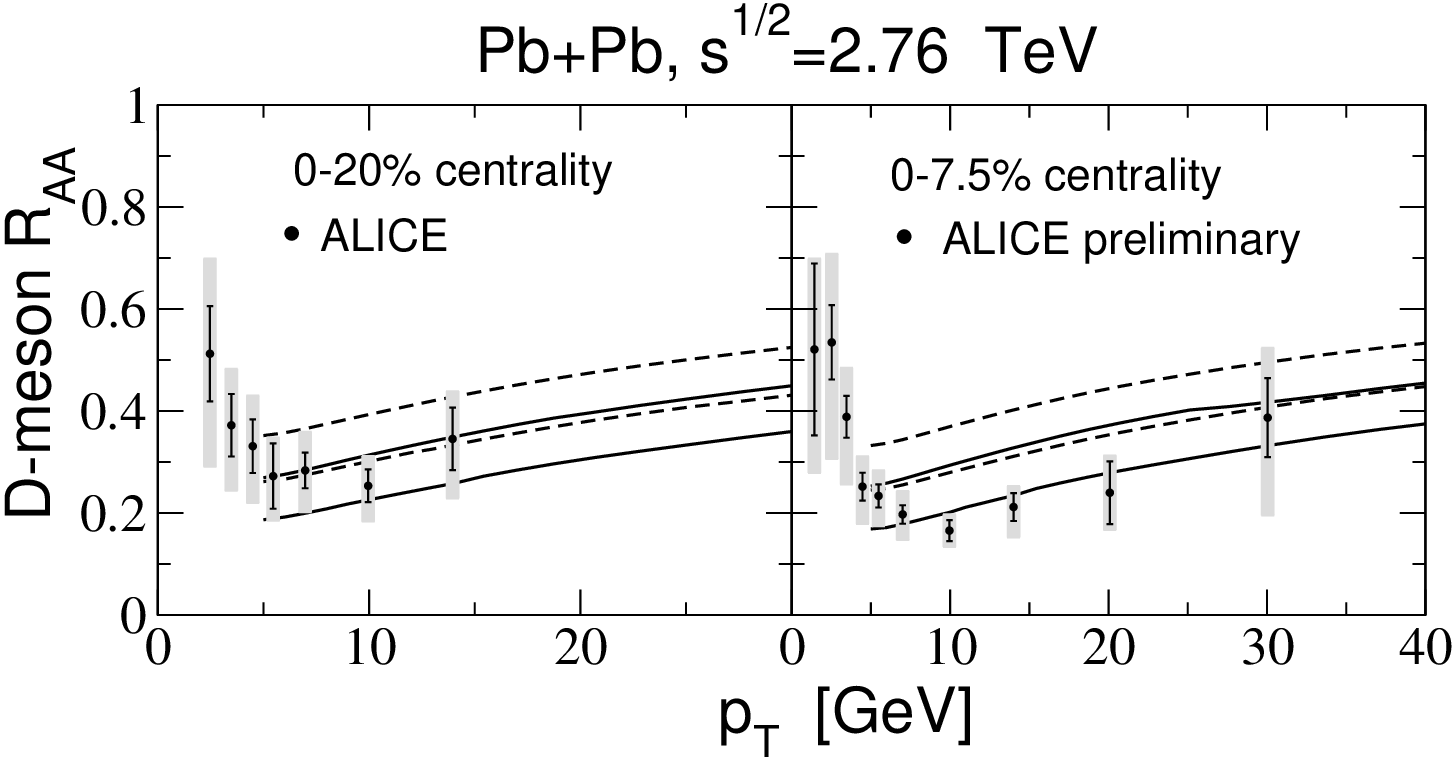}
\caption[.]{
$R_{AA}$ of $D$-mesons
for 0--20\% (left) and 0--7.5\% (right) central 
Pb+Pb collisions at $\sqrt{s}=2.76$ TeV for $\alpha_{s}^{fr}=0.4$
(upper curves) and 0.5 (lower curves).
The solid line: radiative and collisional energy loss.
The dashed line:
purely radiative mechanism.
Data points are from ALICE \cite{ALICE_RAA_D1} (left),
\cite{ALICE_RAA_D2} (right).
Systematic errors
are shown as shaded areas. 
}
\end{wrapfigure}
The tendency of the decrease of $\alpha_{s}^{fr}$ from RHIC to LHC, 
observed first in \cite{Z_RHIC-ALICE}, 
is natural, since 
the thermal reduction of $\alpha_{s}$ should be stronger at the LHC energies.
Thus, the values $\alpha_{s}^{fr}=0.5$ and $0.4$ seem to 
be reasonable benchmarks for calculations of nuclear suppression
for heavy flavors at RHIC and LHC energies.

In Fig.~2 we compare results of our model
with STAR \cite{STAR_e} and PHENIX \cite{PHENIX2_e}
data on the electron $R_{AA}$.
In Fig.~2 we show the total (charm plus bottom) $R_{AA}$ with 
and without collisional energy loss.
Comparison to the data from ALICE  \cite{ALICE_e}
is shown in Fig.~3. There we show the total (charm plus bottom) 
and separately charm and bottom 
$R_{AA}$ with collisional energy loss.
Figs.~2,~3 demonstrate that 
the same window of $\alpha_{s}^{fr}$ as for light hadrons
leads to a quite satisfactory agreement with data on the electron $R_{AA}$.
Similarly to data for light hadrons the electron data
support $\alpha_{s}^{fr}\approx 0.5$ for RHIC,
and $\alpha_{s}^{fr}\approx 0.4$ for LHC.
Thus, the simultaneous 
description of the nuclear suppression of light hadrons and single electrons
in the pQCD picture seems quite 
possible.

In Fig.~4 we compare our results 
with the ALICE data 
\cite{ALICE_RAA_D1,ALICE_RAA_D2} on the $R_{AA}$ for $D$-mesons in 
Pb+Pb  collisions 
at $\sqrt{s}=2.76$ TeV for 0--20\% and 0--7.5\% centrality bins.
Fig.~4 shows the results for the $c\to D$ fragmentation.
We have found that the process $b\to B\to D$ increases $R_{AA}$ 
only by about 2\%.
From Fig.~4 we can conclude 
that the same window in $\alpha_{s}^{fr}$ as for
light hadrons allows  to obtain a fairly reasonable description
of the $D$-meson data as well.

\vspace{.2cm}
\noindent {\bf 4}. 
In summary, we have analyzed the 
RHIC and LHC data on $R_{AA}$ for light hadrons, single electrons, and
$D$-mesons.  
We  have found that once $\alpha_{s}$ is fixed from data on $R_{AA}$ for 
light hadrons it gives a satisfactory agreement with
data on the electron and $D$-meson $R_{AA}$ as well. 
Our results give support for the pQCD 
picture of parton energy loss both for light and heavy flavors.

\section*{Acknowledgments}
I am grateful to the organizers for such an enjoyable and stimulating meeting
and for financial support of my participation.

\end{document}